# MOTIVATING GOOD PRACTICES FOR THE CREATION OF CONTIGUOUS AREA CARTOGRAMS

Shi Tingsheng, Ian K. Duncan, Yen-Ning Chang, Michael T. Gastner[*]

Yale-NUS College, 28 College Avenue West, #01-501 Singapore 138533
[*]Corresponding author: michael.gastner@yale-nus.edu.sg

*Abstract*
*Cartograms are maps in which the areas of regions (e.g., countries or provinces) are proportional to a thematic mapping variable (e.g., population or gross domestic product). A cartogram is called contiguous if it keeps geographically adjacent regions connected. Over the past few years, several web tools have been developed for the creation of contiguous cartograms. However, most of these tools do not advise how to use cartograms correctly. To mitigate these shortcomings, we attempt to establish good practices through our recently developed web application go-cart.io: (1) use cartograms to show numeric data that add up to an interpretable total, (2) present a cartogram alongside a conventional map that uses the same color scheme, (3) indicate whether the data for a region are missing, (4) include a legend so that readers can infer the magnitude of the mapping variable, (5) if a cartogram is presented electronically, assist readers with interactive graphics.*

## INTRODUCTION

Visualizing quantitative data associated with geographic regions is a common task in cartography. Conventionally, quantitative data are shown on choropleth, dot-density, or proportional symbol maps (Dent et al., 2008). Typically, these maps are based on standard map projections (e.g., a transverse Mercator or Albers projection) so that the regions' boundaries have shapes that look familiar to most readers. The most salient feature of a region on a map is arguably the size of the area that the region occupies. However, the area of a region on a conventional map is usually not related to the data the map aims to communicate. For example, Vienna occupies only 0.5% of the area in the choropleth map of Austria in Figure 1a, where Vienna is abbreviated as "WI". Even though the color legend indicates that Vienna is densely populated (data from Wikipedia, 2020a), it is not immediately apparent from the choropleth map how large Vienna's population size is in comparison with the population of other regions.

Cartograms promise a way to overcome the limitations of conventional thematic maps (Raisz, 1934; Hennig, 2013). On a cartogram, every region has an area proportional to its numeric data value. For instance, on the cartogram in Figure 1b, Vienna occupies 21.4% of Austria's area, which accurately represents Vienna's share of the nationwide population. In addition to scaling all areas in proportion to population, the cartogram in Figure 1b is also contiguous (i.e., all neighbors on the conventional map projection in Figure 1a are also neighbors in Figure 1b, and vice versa). Several algorithms have been proposed that create contiguous cartograms (e.g., Dougenik et al., 1985; Merrill et al., 1992; Gusein-Zade and Tikunov, 1993; Sun, 2013). The example in Figure 1b was generated with the fast flow-based method by Gastner et al. (2018), but the principles we propose in this paper are not specific to contiguous cartograms produced with this technique.

Contiguous cartograms have often been criticized as difficult to read because they inevitably appear distorted when they are compared to traditional maps (Fotheringham et al., 2007; Eckert et al., 2008; Woodruff, 2008). We argue that the difficulty is partly caused by the way cartograms are often presented. In this paper, we propose a set of guidelines for the presentation of cartograms that should alleviate some of the concerns about their use. We advocate the adoption of these practices to ensure that cartograms are used appropriately and remain accessible to readers. We also describe the cartogram web application *go-cart.io* (Gastner, 2020) that nudges users to adhere to these guidelines.

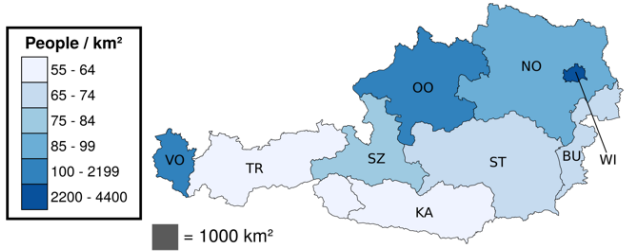 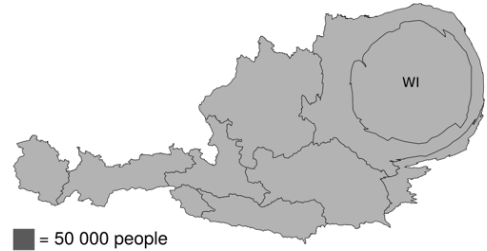

*Figure 1. Two maps that visualize the population distribution of Austria by federal state (Bundesland). (a) Choropleth map of the population density. (b) Cartogram in which areas are proportional to population size.*

**PREVIOUS WORK**

In recent years, several software tools have been developed for the creation of cartograms. Markowska and Korycka-Skorupa (2015) conducted an evaluation of five existing cartogram generation tools: Cartogram Utility for ArcGIS (ESRI, 2007), ScapeToad (Andrieu et al., 2008), OpenGeoDa (Anselin and McCann, 2009), MapViewer (Golden Software, 2010), and MAPresso (Herzog, 2005). They assessed whether these five tools contain certain features that may help users to better read and generate cartograms. Cartogram Utility for ArcGIS and ScapeToad received the highest scores. For example, these cartogram tools were the only two in the list that enabled users to save and export the cartograms as vector graphics. However, not even these two cartogram generators intend to guide users towards proper use of cartograms. None of the available software informs users whether their data are suitable to be shown on a cartogram, how to handle missing data, or how to add meaningful legends.

Clear guidelines for the proper use of cartograms have generally been lacking in the cartogram literature. A noteworthy exception is a user evaluation by Dent (1975). Based on experimental results, he advises to label all administrative units on the cartogram, show a conventional map as an inset in the cartogram, and provide an "anchor stimulus". The intention of the anchor stimulus is to allow map readers to convert between cartogram areas and the numbers that the areas represent. Dent (1975) presents as an example a population cartogram of the northeastern United States. The anchor stimulus in his example takes the form of a square next to the cartogram that is accompanied by the label: "represents 1,500,000 persons". We show a similar anchor stimulus in the cartogram of Austria in Figure 1b.

Since the publication of Dent's (1975) guidelines, cartograms have become a more frequent form of data visualization, but only few cartogram designers have strictly followed his advice. The news media nowadays often adopt cartograms as visualizations accompanying their online content. Some of these cartograms indeed contain an inset map (Los Angeles Times, 2012; New York Times, 2012) or an anchor stimulus (Fairfield, 2008), but the use of these features is not universal. It is understandable that there has been no uniform approach to presenting cartograms so far because their production has required manual adjustments specific to each individual data set. Because of the increasing interest in cartograms, it is timely to implement Dent's (1975) guidelines as an integral part of software that can automate many of the steps needed to convert raw data into a cartogram. A fresh look at the problem also presents an opportunity to establish best practices for cartograms that are presented on a computer screen rather than on printed paper.

When cartograms are shown in a web browser, we can take advantage of modern web technologies (e.g., D3.js or AJAX) to add new elements of interactivity that have the potential to make cartograms more understandable and more enjoyable to users. On news portals, there are already many online cartograms that allow readers to explore the data interactively (Byron et al., 2008; Clark and Houston, 2012; Los Angeles Times, 2012; New York Times, 2012; The Telegraph, 2015; Franklin et al., 2016; BBC News, 2019; Kommenda et al., 2019). There are three categories of interactive features that either have been implemented on these websites or have been proposed in the cartogram literature: animations, linked brushing, and infotips. Animations can help visualize how one map morphs into another. Ware (1998) strongly advocated the use of animations that show a continuous transformation from a conventional map to a cartogram and vice versa. In an experiment, she found that participants with access to animations were more likely to give correct answers, though at the expense of a longer response time. While an animation only displays a single map at a time, linked brushing enables users to compare a cartogram shown alongside a corresponding conventional map. Linked brushing is a mouse-hover effect where the user pointing at one region on the conventional map causes the corresponding region on the cartogram to be highlighted (e.g., by changing its color) and vice versa. Dykes (1997), Haining (2003), and Tobler (2004) all proposed linked brushing as a method of visually communicating the thematic connection between the conventional map and the cartogram. Another mouse-hover effect is an infotip that appears as a pop-up with data about the region under the mouse pointer (Figure 2). Nusrat et al. (2018) proposed infotips as a way to reveal the exact value of the numeric data represented by the region's area. Infotips have been implemented in cartogram software such as GeoViz Toolkit (Hardisty and Robinson, 2011) and MAPresso (Herzog, 2005). However, these software tools are no longer up to date. The cartogram feature of GeoViz Toolkit has been reported to be broken[1], and MAPresso is no longer supported by contemporary web browsers because it is a Java Applet.

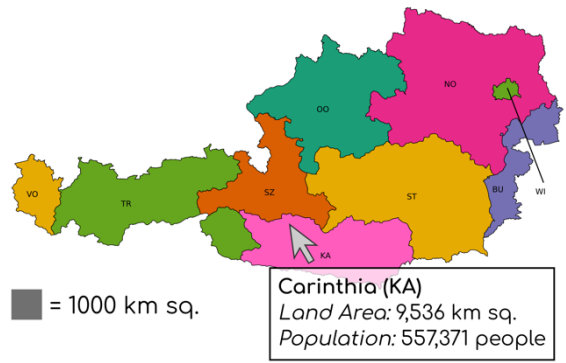 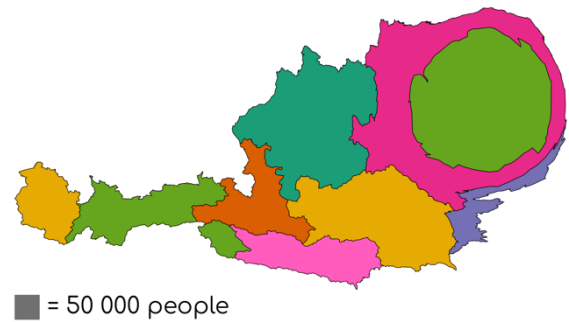

*Figure 2. Screenshot of the go-cart.io user interface. A conventional equal-area map of Austria (left) and a population cartogram (right) are shown alongside each other.*

## GO-CART.IO

We have developed the web application *go-cart.io* (Gastner, 2020) as a free alternative to existing cartogram software that takes advantage of state-of-the-art web technology. It aims to provide an intuitive interface to create cartograms supported by backend computer code that outputs the desired cartograms within seconds (Gastner et al., 2018; Tingsheng et al., 2019). By design, *go-cart.io* neither requires technical knowledge about programming nor access to geographic information system software so that even non-experts can create cartograms from their own data. To generate a new cartogram with *go-cart.io*, users first select a desired input map (e.g., a map of Austria). By default, the website displays a conventional equal-area projection of the chosen map region alongside a population cartogram (Figure 2). Users can fill in their own numeric data for each administrative region either by typing directly in the browser or by uploading a CSV spreadsheet. The website then visualizes the input as a pie chart and as a cartogram. By selecting different data sets from a drop-down menu, users can compare different cartograms (Figure 3). Users can explore the maps with interactive features—animations, linked brushing, and infotips—implemented with the D3.js library. With a simple mouse click, users can download the maps in GeoJSON or SVG vector graphics format. A tutorial with an explanation of the cartogram generation process is available on *go-cart.io*.

At this point in time, the development of *go-cart.io* is still work in progress. While working on the software, we had to make many design decisions that, in our opinion, set an example for the appropriate use of cartograms. The purpose of this paper is to summarize our design decisions as guidelines that we regard as essential if cartograms are to become an effective form of data visualization.

## RECOMMENDED PRACTICES

### Interpretable Total

The main purpose of a cartogram is to show the share of a region's importance in the context of a larger geographic entity. For instance, we may want to show the proportion of Austria's GDP that originates in each federal state (Bundesland). We can visualize these numbers on a cartogram where each Austrian state has an area in proportion to its GDP (data from Wikipedia, 2020b). On such a cartogram, the states' GDPs add up to an interpretable total: Austria's national GDP.

---
[1] https://code.google.com/archive/p/geoviz/issues/39, accessed on 2020-24-05.

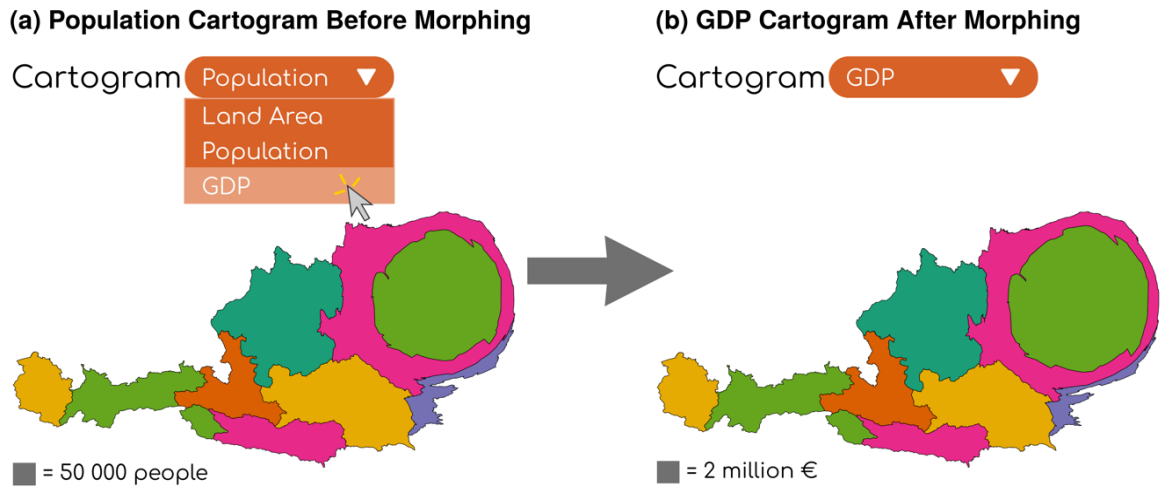

*Figure 3. Users can select a cartogram from a drop-down menu. In this example, the user previously uploaded GDP data. Upon selecting "GDP", the cartogram continuously morphs from a cartogram where areas are proportional to population to a new cartogram where areas are proportional to GDP.*

Unfortunately, many cartograms have been presented in the literature where cartogram regions are proportional to non-additive statistics. The most common mistake is the use of per-capita statistics for the regions' areas. For example, it is not appropriate to make a cartogram of Austria where each state has an area in proportion to the state GDP *per capita*. The states' per-capita GDPs do not add up to an interpretable total, so such a per-capita cartogram would not reveal the contribution of a state to the federal GDP. Examples of cartograms incorrectly based on non-additive statistics can be found in Rittschof et al. (1996; books read per capita), Sui and Holt (2008; mortality rate instead of number of fatalities), and Pleger et al. (2014; citations per publication instead of total citations), to name only a few.

To understand why non-additive statistics are unsuitable for cartograms, it helps to draw an analogy with a more common form of data visualization: pie charts. The numbers associated with the slices of the pie must add up to a number that we can attribute to the whole pie. For example, if each slice represents the GDP of one state, the entire pie represents the national GDP—a meaningful quantity that we can show as a label on top of the pie chart. If instead we were to make slices proportional to GDP *per capita*, we would not be able to describe the sum of these numbers with a simple label. The analogy between cartograms and pie charts was first made explicit by the Worldmapper team (Dorling et al., 2007). We find this analogy intuitive and easy to communicate to non-experts. We have, therefore, decided to integrate the pie chart analogy into *go-cart.io*'s cartogram generation process.

Before a cartogram is shown to the user, we first present the input data as a pie chart (Figure 4). Above the pie chart, we state the sum of the input numbers and pose the question: "Is this a meaningful quantity?" The cartogram is only computed after the user confirms that the total number is interpretable. We hope that the pie chart and the accompanying text nudge users to think about whether their data are suitable and, if not, to amend their input. Figure 4a shows an example where the input numbers add up to a meaningful total, namely the total GDP of Austria. The example in Figure 4b illustrates what happens if the input is unsuitable (GDP per capita): the total of 384,300 € is clearly too small to be the total GDP of Austria and too large to be the nationwide GDP per capita.

**Color Scheme**

While Dent (1975) already advised to show conventional map and cartogram alongside each other, he gave no rule how to choose colors on the two maps. Most cartograms are univariate maps: they represent only one variable (e.g., population in Figure 3a or GDP in Figure 3b), and the area is the only visual stimulus that represents this variable. In univariate cartograms, we are free to choose the colors for each region so that regions are easy to identify. We color the regions such that no neighboring regions share the same color, but corresponding regions on the juxtaposed conventional map and cartogram have the same color (see Figure 2 for an example).

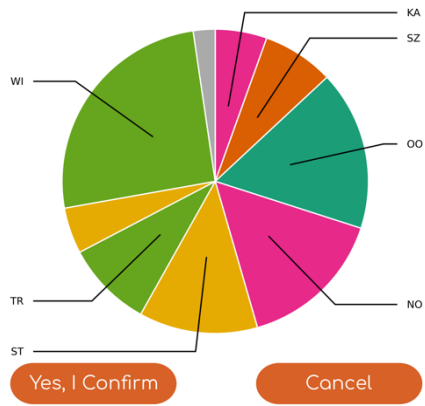
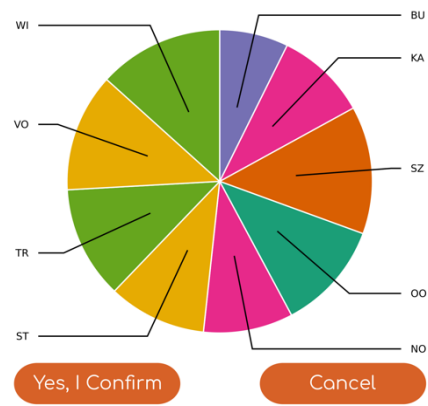

(a) Austria, GDP    (b) Austria, GDP per Capita

*Figure 4. Examples of pie charts shown by go-cart.io during the cartogram generation process.*

By default, *go-cart.io* uses the six-class Dark2 palette from ColorBrewer (Brewer, 2002; Figure 5), which is designed to work well on liquid-crystal displays. Although we believe that the default colors work well in most cases, we allow users to input other color choices, either as Hex codes or with a web-based color picker. The freedom to choose colors may in fact be necessary when representing bivariate data on a cartogram. The website ensures that corresponding regions of the conventional map and the cartogram still share the same color.

## Handling of Missing Data

Real-world geospatial data sometimes contain missing entries for some regions. Conventional thematic maps usually reserve a special color to indicate missing values. We apply the same method in *go-cart.io*. In the default color scheme, we choose light gray for missing data because it clearly stands apart from all other colors in the palette (Figure 5).

Besides setting a special color, we must also specify how large the area of a region with missing data should be on a cartogram. At first glance, it might appear natural to set the area equal to zero. In practice, however, this choice would be a bad solution. If a region's area is equal to zero, it completely disappears from the cartogram, which makes it difficult to compare the cartogram to a conventional map. An area of zero would also incorrectly imply that the statistical variable is known to be zero. A more appropriate choice is to symbolize the absence of knowledge on the cartogram by keeping the region's area the same as on the conventional reference map.

In Figure 6, we illustrate with an example how to represent missing data on a cartogram. On the left, we show a conventional equal-area map of Austria as a reference. On the right, we show a cartogram where the states' areas are proportional to the number of workers in day nurseries. The number for Vienna (WI) is unknown according to the data available from Statistik Austria (2020) We indicate the missing data with a gray fill color. We also keep the area of Vienna in the cartogram the same as Vienna's area in the equal-area map on the left.

## Value-to-Area Legend

Without any additional labels or legends, a cartogram shows the importance of the administrative units only in relative terms. For example, the viewer can judge from the areas of the Austrian states in Figure 1b that Vienna's population is approximately one fifth of Austria's total population. This knowledge is valuable, but users can gain even more insight if they know what magnitude each area represents in absolute terms. Therefore, we regard it as good practice to include a value-to-area legend in every cartogram.

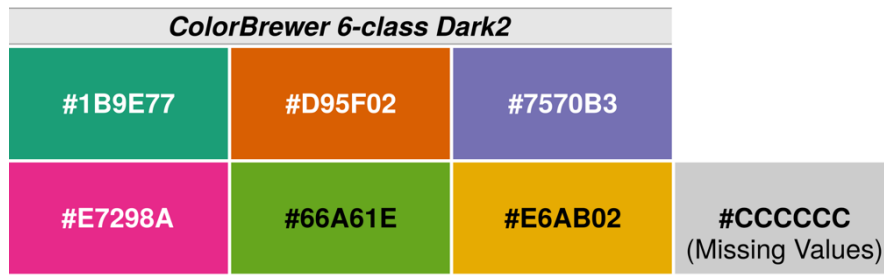

*Figure 5. Six-class color scheme used for cartograms in go-cart.io. The text strings are the Hex codes of each color. Gray is reserved for missing values.*

Dent (1975) recommends a square as the legend symbol with an "anchor stimulus" (i.e., area in the square) "at the low end of the value range". Unfortunately, he does not give an explanation how to obtain the area of the square key for a given data set. In his paper, he shows a sample cartogram of the northeastern United States, where the key represents approximately 1.1% of the total population (1.5 million inhabitants out of a total of 132.0 million). We have chosen slightly smaller keys, ranging from 0.5% for the GDP cartogram in Figure 3 to 0.9% for the day-nursery-personnel cartogram in Figure 6b. On a computer screen, the square is approximately 30×30 pixels. The exact percentage that is represented by the key depends on the aspect ratio of the cartogram. We restrict the keys to be "nice numbers" (sensu Wilkinson, 2013; i.e., powers of 10 multiplied by 1, 2, or 5). If users provide a unit as part of their input data, *go-cart.io* inserts the unit as part of the legend string (e.g. € in Figure 3).

In our current implementation, the legend is always placed below the map. The color of the legend is dark gray (Hex code #707070) to avoid confusion with any of the six colors in the Dark2 color scheme and the light gray used for missing data (Figure 5). In future versions, we may allow users to interactively change the legend color and position.

## Interactive Graphics

We integrate three interactive features into *go-cart.io* to render more readable and understandable cartograms. These interactive features are cartogram-switching animations, linked brushing, and infotips.

- **Cartogram-switching animations:** Users can choose between different data sets from a drop-down menu on the screen, as shown in Figure 3. When the user selects a data set, the previously displayed cartogram morphs into a new cartogram that visualizes the selected data. In *go-cart.io*, the transition from one cartogram to another is implemented as a smooth shift of the polygon vertices from start to end position during a one-second time interval. Figure 3 shows the start and end point of the animation when the user chooses to view the GDP cartogram whereas the original cartogram displayed is the population cartogram.

- **Linked brushing:** The cartogram is always displayed alongside a conventional map in *go-cart.io*. As the user hovers the mouse over an administrative region on one of these two maps, the corresponding region is simultaneously highlighted on the other map. In our current implementation, we highlight the region by increasing the brightness of its fill color. Other forms of highlighting (e.g., changing the border color or thickness) are also conceivable and may be implemented in future versions. We demonstrate linked brushing in Figure 2, where we show a population cartogram of Austria alongside a conventional equal-area map. Because the mouse is hovering over Carinthia on the equal-area map, the fill color has become a lighter shade of purple (#FF5CBD) than the default (#E7298A, see Figure 5) on both the equal-area map and the cartogram.

- **Infotips:** As the user hovers the mouse over an administrative region, a pop-up automatically appears at the position of the pointer. The text in the pop-up consists of the region's name and the data (e.g., population and GDP) represented by the areas in the juxtaposed conventional map and cartogram. An example is shown in Figure 2.

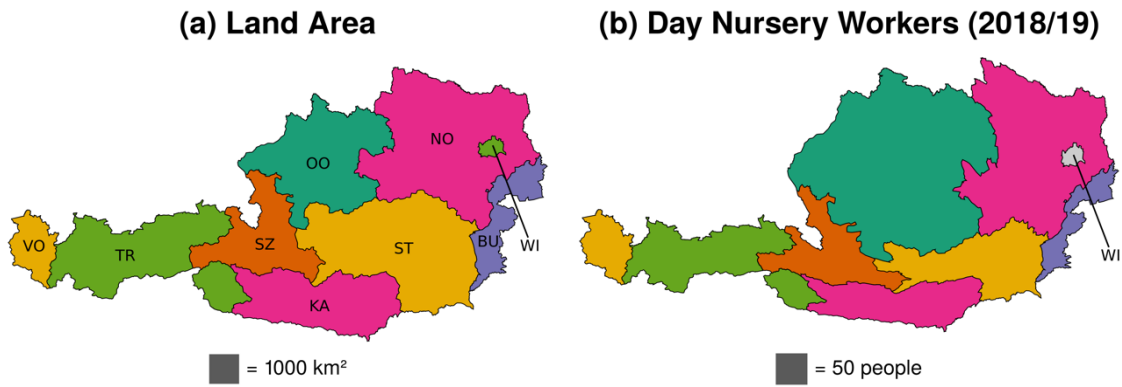

*Figure 6. Maps of Austria. (a) Conventional equal-area map. (b) Cartogram where the areas of the states are proportional to the number of day nursery workers. The number for Vienna (WI) is unknown. We symbolize the missing data with a gray fill color (Figure 5) and with an area that is equal on both maps.*

**DISCUSSION**

The effectiveness of cartograms has been under discussion for the past few decades. In previous literature, cartograms have often been characterized as confusing and difficult to read (e.g., Dent, 1975; Rittschof et al., 1996; Fotheringham et al., 2007; Woodruff, 2008). We acknowledge that cartograms are inherently distorted, but we believe that the previous criticism of cartograms has often been triggered by poor usage rather than a fundamental flaw in the concept of a cartogram. Previous experiments sometimes presented cartograms to the participants without showing a conventional map as a reference (Sun and Li, 2010; Nusrat et al., 2018), which increases the perceived difficulty of understanding cartograms. If there is no value-to-area legend, interpreting a cartogram becomes even more challenging because it is unclear to the reader how to translate the areas to concrete numbers. Moreover, cartograms in the previous literature often translated non-additive statistics into region areas (e.g., Rittschof et al., 1996; Sui and Holt, 2008; Pleger et al., 2014), which renders cartograms almost uninterpretable. However, there is evidence that cartograms have great potential if they are used appropriately. For example, when Dent (1975) tested cartograms based on additive statistics with a value-to-area legend and a conventional map as inset, users described them as "exhibiting a definite style", "interesting", and "innovative".

At the time of Dent's (1975) experiment, readers encountered maps mostly on printed paper. With the rapid development of map-making technologies, cartograms nowadays appear more often on a computer screen than in print. If presented electronically, we can convey information with interactive features that are not possible on printed paper. Three interactive features have been described in the literature: cartogram-switching animations (Ware, 1998), linked brushing (Dykes, 1997; Haining, 2003; Tobler, 2004), and infotips (Nusrat et al., 2018). While other interactive features exist in the general literature on data visualization or cartography (e.g., zooming, panning), they would not add obvious value to cartograms. We have recently completed an experiment in which we tested how these interactive features affect participants' performance in cartogram reading tasks. Our initial results suggest that animations are especially helpful in synoptic tasks (i.e., tasks in which users must summarize large-scale information about the cartogram). For elementary tasks, participants performed well without any interactivity, but the combination of the three investigated interactive features never had any negative impact. Therefore, we recommend the full suite of features (i.e., animations, linked brushing, and infotips) to be available whenever cartograms are presented on a computer screen. All three features are automatically implemented if cartograms are generated with *go-cart.io*.

We are unaware of any experiment that has specifically tested the influence of our other guidelines (i.e., interpretable total, combining conventional map and cartogram with matching colors, indicating missing values, and value-to-area legends). To judge how well the recommended practices integrate as a whole in *go-cart.io*, we conducted a pilot study with 13 college students in Singapore. We used Brooke's (1996) System Usability Scale, which is a widely used tool for the evaluation of user interfaces in industry. In our pilot study, *go-cart.io* received an average score of 70.45 out of 100. This score is close to the median 70.91 for a large sample of System Usability Scale scores collected by Bangor et al. (2008).

Considering that *go-cart.io* is still work in progress, this result is encouraging. Still, more work needs to be done to ensure an even more satisfactory user experience. Currently, administrative units are only labeled on the conventional map, not on the cartogram. Automatic label placement is a complex problem (Freeman, 2007). We are not aware of an existing algorithmic approach to label placement that we can easily adopt for our needs. We also acknowledge that the current version of *go-cart.io* is not ideally suited to making bivariate cartograms. Users who are interested in bivariate cartograms must currently change the regions' colors manually and add a legend with other software.

## ACKNOWLEDGMENTS

We thank Bernard Boey Khai Chen for help with implementing the algorithm to handle missing data. This work was supported by the Singapore Ministry of Education (grant R-607-000-401-114) and a Yale-NUS College start-up grant (R-607-263-043-121).

## BIOGRAPHY

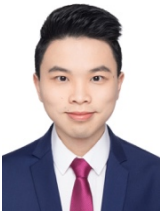

Shi Tingsheng is a student at Yale-NUS College in Singapore. He is pursuing a B.Sc. in Mathematical, Computational, and Statistical Sciences, and will graduate in 2020.

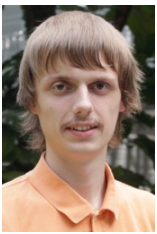

Ian K. Duncan is a student at Yale-NUS College in Singapore. He is pursuing a B.Sc. in Mathematical, Computational, and Statistical Sciences with a minor in Philosophy, and will graduate in 2021.

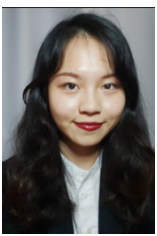

Chang, Yen-Ning is pursuing a B.Sc. in Mathematical, Computational, and Statistical Sciences major at Yale-NUS College. She will graduate in 2021.

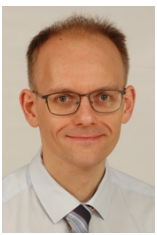

Michael T. Gastner is Assistant Professor for Mathematics, Computer Science and Statistics at Yale-NUS College in Singapore. He received the PhD in Physics from the University of Michigan and held postdoctoral positions at the Santa Fe Institute, the University of Oldenburg, and Imperial College London. Before moving to Singapore, he was a faculty member at the University of Bristol and the Hungarian Academy of Sciences.